# Ultra-open Ventilated Metamaterial Absorbers for Sound-silencing Applications in Environment with Free Air Flows

Xiao Xiang[1], Xiaoxiao Wu[1,2,*], Xin Li[1], Peng Wu[1], Hong He[1], Qianjin Mu[1], Shuxia Wang[1,*], Yingzhou Huang[1], Weijia Wen[2,3,*]

[1]*Chongqing Key Laboratory of Soft Condensed Matter Physics and Smart Materials, College of Physics, Chongqing University, Chongqing 400044, China*

[2]*Department of Physics, The Hong Kong University of Science and Technology, Clear Water Bay, Kowloon, Hong Kong, China*

[3]*Materials Genome Institute, Shanghai University, Shanghai 200444, China*

*Corresponding Authors. E-mails: phwen@ust.hk (W. Wen), wangshuxia@cqu.edu.cn (S. Wang), xwuan@connect.ust.hk (X. Wu).

**Abstract**

High-efficiency absorption of low-frequency sounds (< 1000 Hz) while maintaining a free flow of fluids remains a significant challenge in acoustical engineering due to the rigid trade-off between absorption and ventilation performances. Although ongoing advances in acoustic metamaterials have unlocked unprecedented possibilities and various metamaterial absorbers have been proposed, most of them only work adequately in the condition of no sound transmissions. Unfortunately, such condition requires a complete block of fluid channels due to longitudinal nature of sounds, which allows them to penetrate any small holes. Otherwise, their absorption performance could be drastically degraded and often cannot exceed 50%. This basic trade-off between absorption and ventilation performances definitely constrains their applications in daily scenarios where free air flows are necessary. Though some ventilated sound barriers with large transmission loss have been demonstrated, they essentially only reflect sounds, which are still there and even may be reflected back. Here, to overcome this general difficulty, we propose and demonstrate an ultra-open ventilated metamaterial absorber. The absorber, aiming at low-frequency sounds, simultaneously ensures high-performance absorption and ventilation, confirmed in experiments. Their mechanism is understood from an effective model of coupled lossy oscillators. Furthermore, the absorbers can be simply stacked to work in a customized broadband, while maintaining a good ventilation. The demonstrated absorber provides a clear scheme for achieving high-performance absorption and ventilation at low frequencies, necessary for applications in environment with free air flows.

## 1. Introduction

In the last two decades, various acoustic metamaterial absorbers have been proposed to overcome the intrinsic limits of natural sound absorbing materials, when dealing with low-frequency sounds (<1000 Hz) [1-4]. Once transmissions are blocked, they give rise to anomalous high-efficiency absorptions at customized working frequencies [5-14]. Compared with conventional porous materials, they have compact profiles and can be deployed in harsh environments such as humid and narrow spaces to control the noise and improve the acoustic environment. However, in daily life and industry, the generation of noise is usually associated with the instabilities of background flows, especially the turbulences in or around ducts, nozzles and turbines [15-19]. Moreover, the background flows, such as air or water, must have a free pass for the proper functioning of corresponding devices containing these structures. Such practical scenarios render many previous metamaterial absorbers incompetent, because they only work adequately after flow channels completely sealed to eliminate transmissions since sound can penetrate any small holes [1-4]. Otherwise, if there is a transmission channel, the absorption performance of these metamaterial absorber will be drastically degraded and often cannot exceed 50% [20-23]. Recently, several ventilated metamaterial absorbers have been demonstrated [23-31]. However, their performances, including absorption and/or ventilation, are still not satisfying, as few of them can simultaneously achieve high-efficiency sound absorption (>90%) and ventilation (>60% wind velocity ratio). Their insufficiencies can be roughly classified into three categories: they are only experimentally confirmed at high frequencies (> 1000 Hz) [24,31]; they only have small open area ratios [23,25,27], which cannot bring good ventilation performance; or their absorption is not efficient enough (< 90%) in experiments [26,30]. In contrast, many ventilated sound barriers [32-42] can simultaneously achieve high-efficiency sound reflection (>90%) and ventilation (>60% wind velocity ratio) at low frequencies in experiments. This performance deficiency of ventilated metamaterial absorbers can be ascribed to the fact that the maximum absorption cross-section of a single subwavelength scatter is only one quarter of its maximum scattering cross-section [31,43]. Accordingly, the absorption of a ventilated acoustic metamaterial composed of subwavelength scatters is usually much smaller compared with its reflection. Despite the odds, is it still possible to break the limit and finally achieve a broadband absorber with simultaneously good absorption and ventilation at low frequencies? If so, the designed absorber would have many important applications in noise control of ducts, nozzles, and turbines, which are common in buildings and automobiles. Many stunning structures, such as silencer windows, will also become possible, and can provide good daylight, fresh air, and a silent acoustic environment.

In this work, an ultra-open ventilated metamaterial absorber (UVMA) is proposed and experimentally demonstrated (see also Movies S1-S5 in Supplementary Material), which can simultaneously achieve high-performance acoustic absorption (>95%) and ventilation (>80% wind velocity ratio). The UVMA unit is comprised of weakly coupled split-tube resonators, and its absorption is demonstrated through numerical calculations and experimental measurement. Broadband absorption is then achieved by optimized stacking of UVMA units with different resonance frequencies. Thus, the demonstrated metamaterial absorber provides a direct route for simultaneously achieving high-efficiency absorption and ventilation at low frequencies and may find practical applications in noise-control engineering of an environment filled with flowing fluids such as air conditioners, exhaust hoods, and flow ducts.

## 2. Results and Discussion

### 2.1. Structure of UVMA Metamaterial Absorber

An array of the UVMA units are assembled as a frame constituting the designed metamaterial absorber, as schematically depicted in Fig. 1(a). The particularly large hollows in the frame permit fluid flows, such as air or water, freely passing through the structure. In the following study, it is assumed that the structure is immersed in air. To simultaneously realize efficient sound absorption and air-flow ventilation, the sound wave incident on the UVMA units should be absorbed near perfectly. As depicted, the UVMA units are packaged in a rectangular lattice, and the lattice constants along the *x* and *y* directions are $L$ and $L/4$, respectively. A supercell comprising of four UVMA units is depicted in Fig. 1(b), and the details of a single UVMA unit are shown in Fig. 1(c). The cover is removed to demonstrate the details of the UMVA unit (note that the structure is rotated by 90° for better visualization). Each UVMA is composed of two split-tube resonators, placed symmetrically and coupled through a narrow slit between them [6,23]. The sectional diagram on the *xz*-plane for a single UVMA unit is presented in Fig. 1(d), demonstrating the identical but oppositely oriented split-tube resonators. Aiming at low frequencies, the proper geometric parameters of the UVMA units are determined, such that the units should have optimized absorption and ventilation performance, and we summarize possible design strategies in Section 2.2. Experimental verifications of their absorption performance are conducted with the setup shown in Fig. 1(e) (see Appendix G.3 for setup details), and the results will be discussed in Sections 2.3 and 2.4. Their ventilation performance is also experimentally measured (see Appendix G.4 for setup details), and will be discussed in Section 2.5.

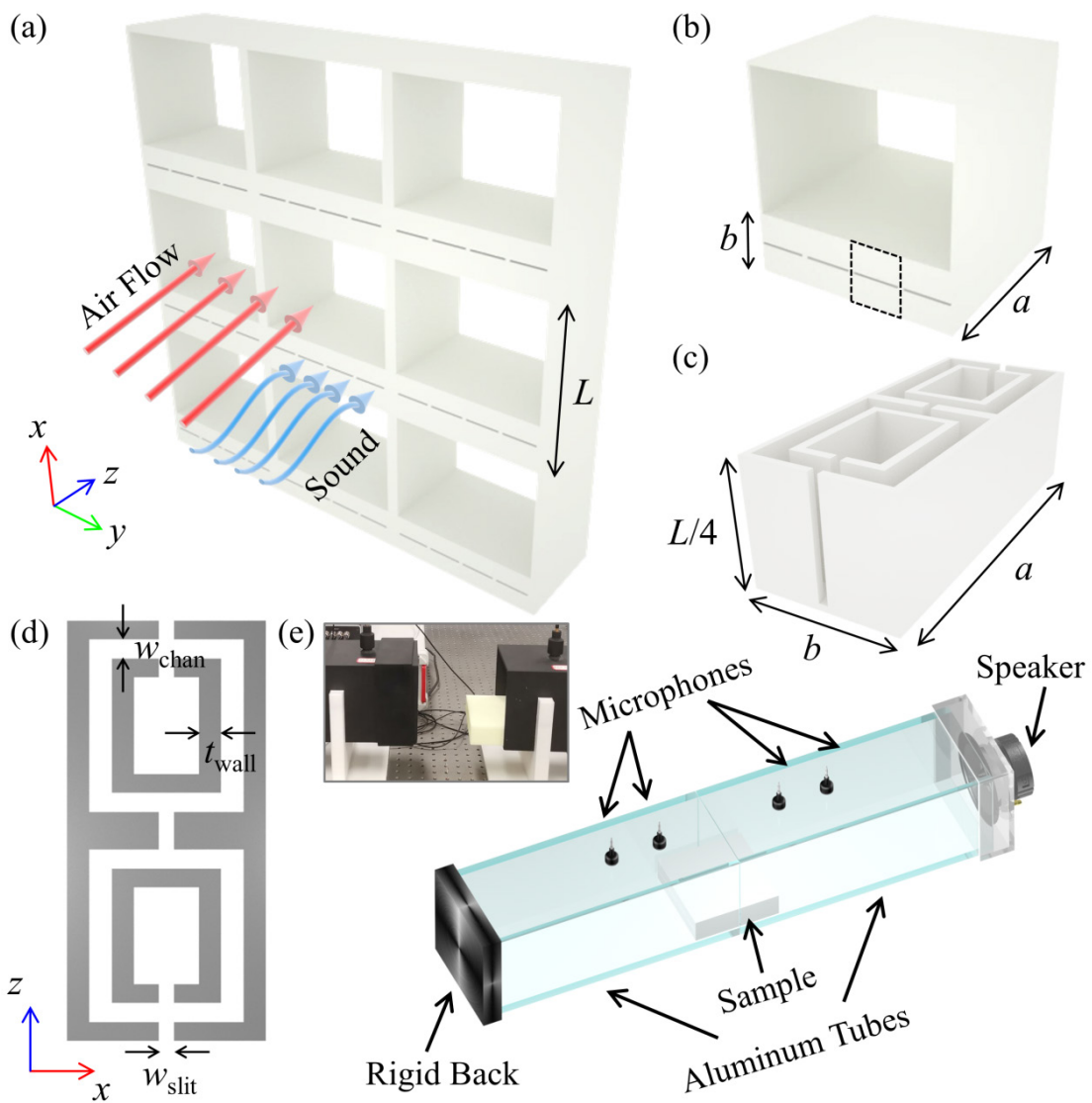

Fig. 1. (a) Perspective schematic of the UVMA units arranged in a rectangular lattice. The lattice constant along $x$ ($y$) direction is $L$ ($L/4$). (b) Close-up view of a supercell. It is comprised of four UVMA units. (c) Perspective view of a single UVMA, as denoted by the dashed rectangle in (b). To demonstrate the details inside, the structure is rotated and the cover is removed. (d) Sectional schematic of the UVMA on the *xz*-plane. (e) Experimental setup for acoustic measurement. The impedance tube has a square cross-section (147×147 mm$^2$), and the standard four-microphone method is adopted. The inset shows the photograph of a fabricated sample placed in the impedance tube.

**2.2. Numerical Study of UVMA Design Strategies**

To investigate the impact of geometric parameters on the acoustic absorption performance of the UVMA, full-wave numerical simulations are performed (see Appendix G.2 for setup details). Complex transmission coefficient *t* and reflection coefficient *r* are retrieved, and the absorption coefficient *A* is calculated as $A = 1 - |t|^2 - |r|^2$, due to the conservation of energy. The key geometric parameters (denoted in Fig. 1(b)–(d)) are the length *a*, height *b*, width of the channels $w_{\text{chan}}$, and width of the slits $w_{\text{slit}}$, which effectively alter the acoustic performance of the UVMA.

First, the length *a* is considered while fixing the other parameters ($b = 40$ mm, $w_{\text{chan}} = 1.4$ mm, $w_{\text{slit}} = 1.4$ mm, $t_{\text{wall}} = 2$ mm). The calculated absorption is plotted as a 2D color map of the length *a* and the frequency, as shown in Fig. 2(a). The red strip in the color map highlights the shift of the resonance of the UVMA unit. As the length *a* is increased, the resonance shifts towards lower frequencies, and the absorption reaches near-unity. Next, the height *b* is considered, which controls the UVMA's open area ratio while fixing other geometric parameters. The corresponding absorption is plotted versus $b$ and frequency, as shown in Fig. 2(b). The height $b$ (25–60 mm) considered here corresponds to the open area ratio of 59–83%. Similarly, as the height $b$ increases while the open area ratio decreases, the resonance shifts towards lower frequencies, and the absorption also reaches near-unity.

The cases are more interesting for the widths of the channels $w_{\text{chan}}$ and slits $w_{\text{slit}}$, as shown by the corresponding absorption color maps displayed in Figs. 2(c) and 2(d), respectively. It can be seen that with a decreasing width of the channels $w_{\text{chan}}$ or slits $w_{\text{slit}}$, the acoustic absorption of the UVMA improves significantly, and the resonance also shifts towards lower frequencies. More importantly, when $w_{\text{chan}}$ becomes small enough, the two moderate absorption resonances (~ 60%) will coalesce into a single near-unity peak. Thus, very narrow channels and slits (<2.0 mm) must be employed to ensure an efficient absorption. The coalescence and enhancement of the absorption resonances can be understood using an effective model of coupled lossy oscillators (see Appendix B). Briefly, the key parameters affecting the absorption spectra are the visco-thermal loss $\eta_l$, the radiation loss $\eta_r$, and the radiation coupling $\eta_c$ between the two resonators. It is found that, under the subwavelength assumption, the condition of coalescence is then (see Appendix B)

$$|\eta_c| \leq \frac{\eta_l + \eta_r}{\sqrt{3}}. \tag{1}$$

Therefore, the condition shows that we can maintain the high-efficiency absorption (>80%) of the UVMA units, meanwhile shifting its resonance in a large range. The only requirement is that we keep the narrow channels $w_{chan}$ and slits $w_{slit}$ untouched, as they ensure a large visco-thermal loss $\eta_l$ and a small radiation coupling $\eta_c$, respectively. This fact suggests a possibility of optimizing the absorber for different working frequencies and ventilation conditions, which we will employ in Section 2.4 to design absorbers with customized broadband.

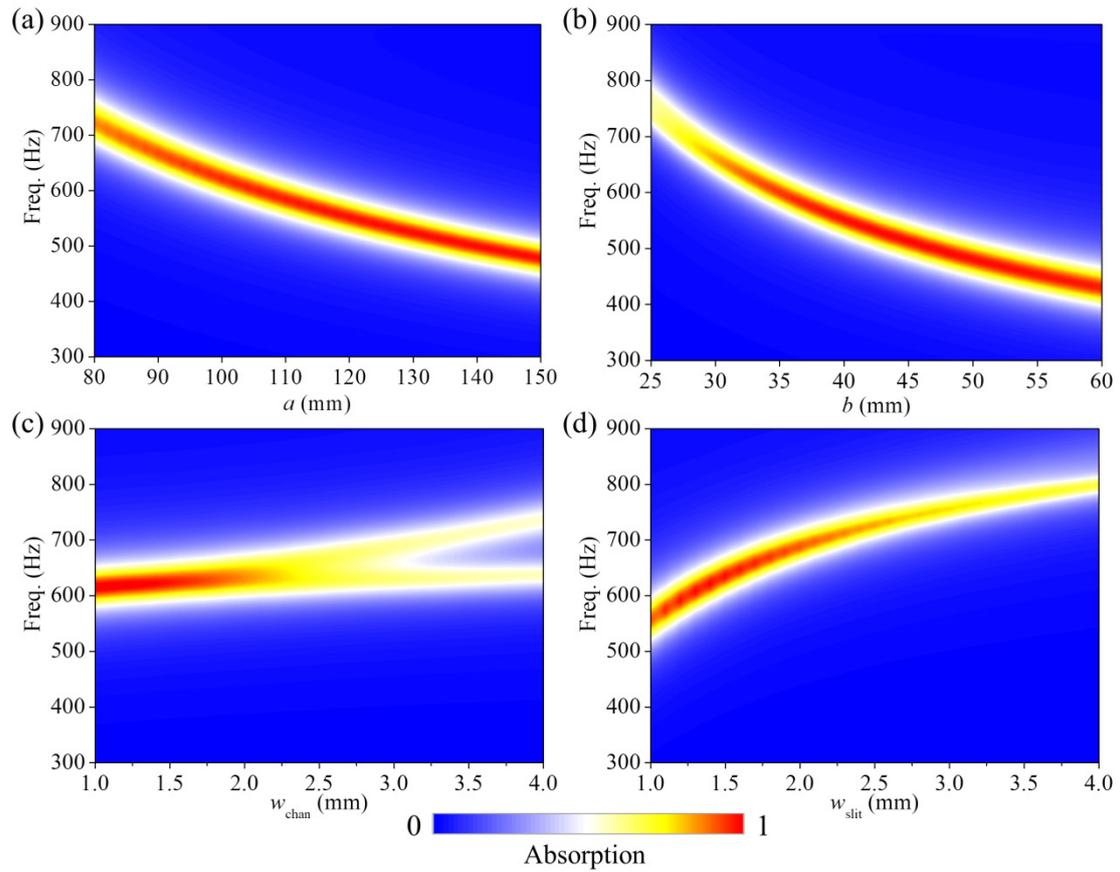

Fig. 2. (a)-(d) Simulated spectra of absorption as the function of frequency and the geometric parameter *a* (a), *b* (b), $w_{chan}$ (c), or $w_{slit}$ (d), respectively. All Start from the configuration *a* = 100 mm, *b* = 40 mm, $w_{chan}$ = 1.4 mm, $w_{slit}$ = 1.4 mm, $t_{wall}$ = 2 mm (when tuning one parameter, the others are kept unchanged). The red strips highlight the shift of resonances when the geometric parameters are tuned.

## 2.3. Experimental Demonstration of Acoustic Performance

Experimental measurement of acoustic properties of the UVMA units are then conducted. Two samples are considered and are labeled as Design I ($a$ = 100 mm, $b$ = 40 mm, $w_{chan}$ = 1.4 mm, $w_{slit}$ = 1.4 mm) and Design II ($a$ = 150 mm, $b$ = 45 mm, $w_{chan}$ = 1.4 mm, $w_{slit}$ = 1.4 mm) (see Appendix G.1 for their fabrication). The open area ratio of Design I (II) is 72.8% (69.4%). The measured transmission and reflection of the two samples (dotted lines) are shown in Fig. 3(a), which agree well with the simulated results (solid lines). Both reflection and transmission spectra exhibit dips near the resonance frequencies, which implies high-efficiency absorptions. As shown in Fig. 3(b), the simulated and measured absorptions demonstrate quantitative agreement between each other. In experiments, for Design I (II), the measured absorption reaches 93.6% (97.3%) at 637 Hz (472 Hz), as indicated by the red (purple) arrow. For reference, the acoustic performances of two melamine foams (Basotect G+, BASF, Germany) [23] are also measured. They are labeled as Foam I and Foam II, which have the same dimensions as Design I and Design II, respectively, and their absorptions are plotted as gray solid lines in Fig. 3(b). The UVMA units clearly demonstrate superior acoustic performances near the resonances compared to the commercial foams. This superiority is more clearly demonstrated if the data are plotted in the dB scale (see Fig. A.1 in Appendix A), which are more industrially relevant [44]. This superiority is also straightforwardly manifested in the demonstration videos (see Movies S1 and S2 in Supplementary Material), where we can directly hear the contrasting absorption performances of Design II and Foam II, respectively.

In order to understand the physical mechanism behind the high-efficiency absorption of the UVMA units at resonances, especially why it can significantly exceed 50%, the simulated cross-sectional acoustic pressure fields of Design I are plotted. The pressure amplitude at the resonance (610 Hz) is shown in Fig. 3(c). Similar to a single split-tube resonator [6,23], the absorption of the UVMA is caused by the friction of oscillating air flows in the long and narrow channels, which dissipates the incident sound energy as heat. Since the air flow is driven by the pressure difference between the inside cavities and the outside environment, the UVMA can offer a significant absorption in the condition of no porous materials, as the pressure difference is largely enhanced at resonance.

Moreover, the acoustic pressure of the two resonators at resonance demonstrates a 90°-phase difference, as shown in Fig. 3(d). This phase difference suggests that the mirror plane ($z = 0$) could be treated as a superposition of an acoustic soft boundary introducing a 180°-phase difference and an acoustic hard boundary introducing 0-phase difference. We have performed simulations involving only a half of the UVMA, terminated by an acoustic soft or hard boundary, respectively. It is found that the absorption of the UVMA is the average of the absorptions of the two situations (see Appendix D), hence confirming that the treatment is exact. As both acoustic hard and soft boundaries act as back reflectors [5-7], they cause multiple scatterings of the incident sound that hybridize resonance modes of the two single split-tube resonators. This hybridization is the key to achieve an effective absorption [20,21]. Further, because the waves reflected by acoustic hard and soft boundaries have a 180°-phase difference, they would tend to cancel each other, ensuring a near-unity absorption [23]. The origin of this 90°-phase difference at resonance can also be understood in the frame of the model of coupled lossy oscillators (see Appendix C). In addition, due to their subwavelength profile, the UVMAs still demonstrate good absorption performance under oblique incidence even for large angles (see Appendix E).

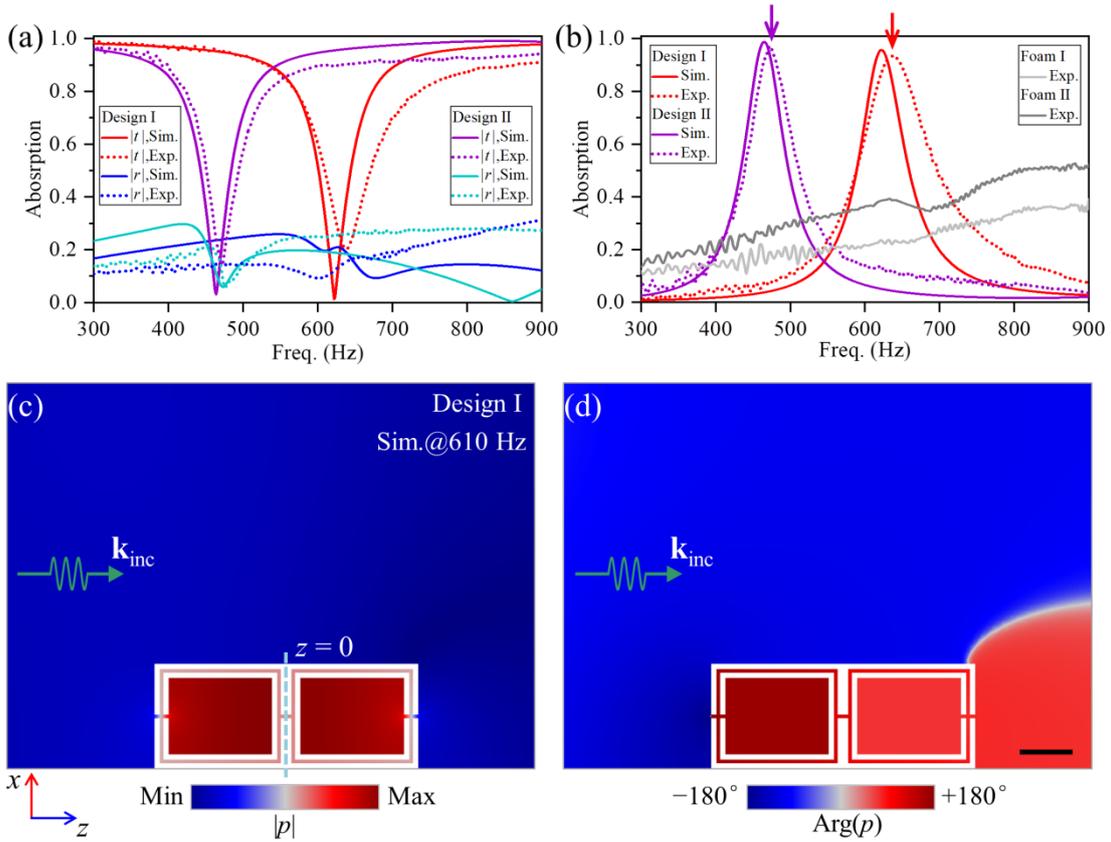

Fig. 3. (a) Simulated (solid lines) and experimentally measured (dashed lines) transmission and reflection spectra of the UVMA units. For Design I (open area ratio 72.8%), $a$ = 100 mm, $b$ = 40 mm, $w_{chan}$ = 1.4 mm, $w_{slit}$ = 1.4 mm. For Design II (open area ratio 69.4%), $a$ = 150 mm, $b$ = 45 mm, $w_{chan}$ = 1.4 mm, $w_{slit}$ = 1.4 mm. (b) Simulated (colored solid lines) and experimentally measured (dotted lines)

absorption spectra of Design I and Design II. Gray solid lines represent measured absorption spectra of melamine foams (Basotect G+, BASF, Germany). As a reference sample, Foam I (II) has the same dimensions as Design I (II). (c),(d) Simulated pressure field maps for Design I on the slice *y* = 0 (central cross section), when the frequency is at resonance (610 Hz). The pressure field in the resonators exhibit a strong enhancement (c), and a 90°-phase difference between the two resonators (d). The green arrows indicate the incident sound. The black scale bar is 20 mm.

### 2.4. Customized Broadband Acoustic Absorption

As discussed in the previous literature [44], the causal nature of an acoustic response imposes a fundamental inequality that relates the two most important aspects in sound absorptions: the absorption spectrum and the absorber length. However, only the no-transmission situation was considered, and what is the limit when transmission is allowed and an open area ratio is specified? It would clearly be complicated for a general case, but if the absorber is mirror symmetric, as aforementioned, the open area can be treated as the superposition of an acoustic soft boundary and an acoustic hard boundary (see Appendix D). Thus, the minimum length $a_{\min}$ of the absorber should be twice the case when the transmission channel is blocked. That is, we have

$$a_{\min} = \frac{1}{2\pi^2} \frac{B_{\text{eff}}}{B_0} | \int_0^\infty \ln[1 - A(\lambda)] \mathrm{d}\lambda |, \qquad (2)$$

where $B_0$ is the bulk modulus of the background fluid (air), $B_{\text{eff}}$ is the effective bulk modulus of the metamaterial in the static limit, $A(\lambda)$ is the absorption spectra, and $\lambda$ is the acoustic wave length [44]. For a broadband absorber comprised by finite number (*N*) of absorbing units, their resonance frequencies should be exponentially spaced for optimal performance [44]. Therefore, the resonance frequencies should be selected as:

$$f_n = f_0 e^{\varphi n}, \tag{3}$$

where, $n = 1$ to $N$, $f_0$ is the cut-off frequency, and $\varphi$ is a coefficient determined by the target frequency band. Broadband UVMA units are constructed in the targeted frequency bands through this relationship. Given the trade-off between absorption and ventilation, and to maintain an efficient ventilation, a broadband UVMA comprised of seven units is considered (shown in Fig. 4(a)). As an example, the frequency band 478−724 Hz is targeted, and the discretized resonance frequencies of the units are selected according to Eq. (3) (see Design III of Table 1). The corresponding lengths *a* of the units are determined by referring to Fig. 2(a). The simulated (solid line) and measured (dotted line) absorption spectra are shown in Fig. 4(b). The working bandwidth is 465−765 Hz in simulations (which is defined as the frequencies where the absorption exceeds 50%), corresponding to a large fractional bandwidth of 48.8%. Meanwhile in the experiments, the working bandwidth is 476−726 Hz, corresponding to a fractional bandwidth of 41.6% [12]. The designed absorber has successfully covered the targeted frequency band and the simulated (measured) average absorption in the band is 73.3% (69.3%). The small discrepancy can be attributed to the fabrication and measurement errors during the experiments in one aspect, and the assumption of infinite acoustic impedance for the solids in simulations in the other aspect. Nevertheless, with an open area ratio of 52.4%, the results confirm that the stacking scheme has led to a broadband UVMA.

There is an evident trade-off between the bandwidth and the average absorption. If a narrower frequency band is targeted instead (for example, 478−620 Hz), the absorption will be enhanced (Fig. 4(c)). With the selected resonance frequencies (see Design IV of Table 1), the simulated (measured) average absorption in the targeted band becomes 85.8% (81.4%). Further, multiple discontinuous frequency bands can be aimed. For instance, considering the dual bands 478−550 Hz and 640−690 Hz, four resonance frequencies are chosen for the first band, and three are chosen for the second band (see Design V of Table 1). The simulated (measured) absorption spectrum is shown in Fig. 4(d) and the average absorption in the dual bands is 83.1% (78.1%). Their corresponding geometric parameters are summarized in Table 2. The performance of the dual-band sample is also directly demonstrated (see Movies S3 in Supplementary Material). The results confirm that a UVMA can be designed with customized working frequency bands, meanwhile permitting the pass of sounds at required frequencies.

Table 1. Resonance frequencies of UVMA units for the samples with customized broadband absorptions. The unit of the frequencies is Hz.

| Design | Unit 1 | Unit 2 | Unit 3 | Unit 4 | Unit 5 | Unit 6 | Unit 7 |
|--------|--------|--------|--------|--------|--------|--------|--------|
| III    | 478    | 512    | 549    | 588    | 630    | 676    | 724    |
| IV     | 478    | 499    | 521    | 544    | 568    | 593    | 620    |
| V      | 478    | 501    | 525    | 550    | 640    | 665    | 690    |

Table 2. Geometric parameter *a* of Designs III, IV, and V. The unit of *a* is mm. In all the three designs, other geometric parameters are *b* = 40 mm, $w_{chan}$ = 1.4 mm, $w_{slit}$ = 1.4 mm.

| Design | Unit 1 | Unit 2 | Unit 3 | Unit 4 | Unit 5 | Unit 6 | Unit 7 |
|--------|--------|--------|--------|--------|--------|--------|--------|
| III    | 75     | 68     | 60     | 54     | 49     | 44     | 40     |
| IV     | 75     | 70     | 65     | 61     | 58     | 53     | 50     |
| V      | 75     | 70     | 65     | 60     | 48     | 45     | 43     |

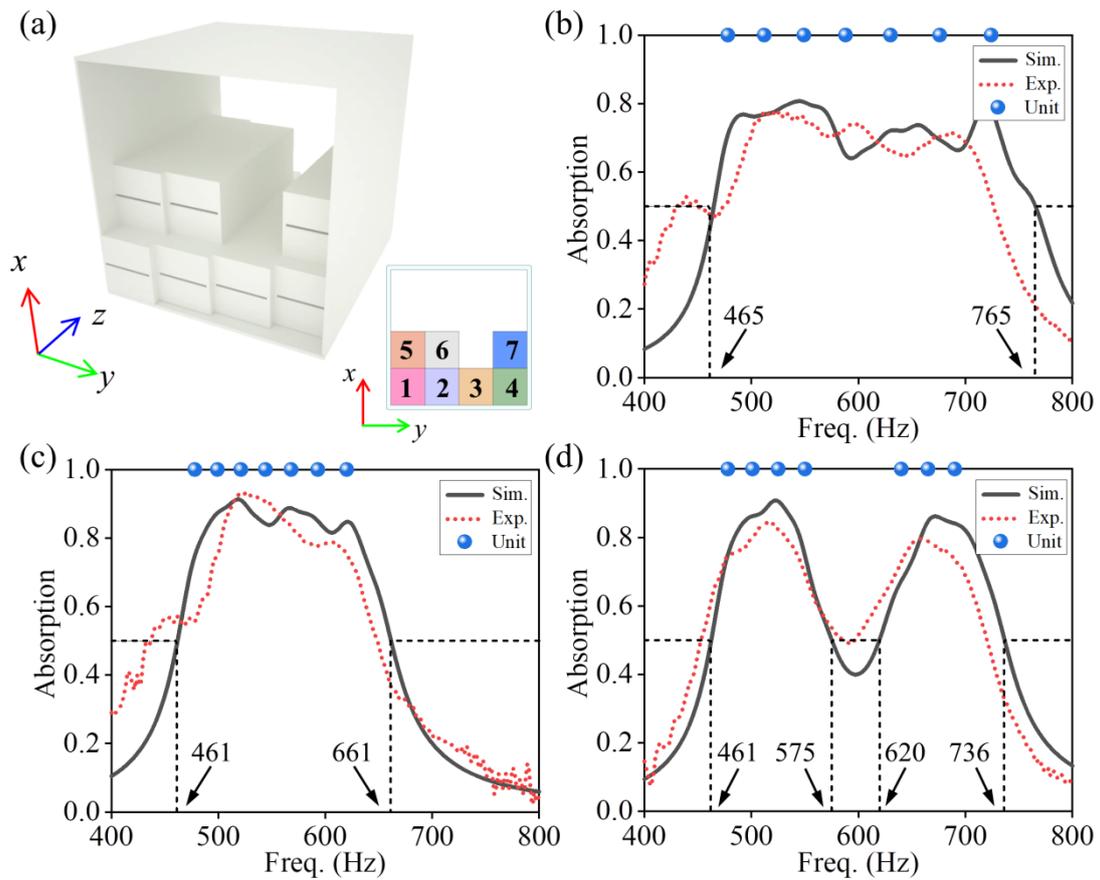

Fig. 4. (a) Perspective schematic of a broadband UVMA comprised of 7 units (open

area ratio 52.4%), with resonance frequencies increasing from units 1 to 7, and their stacking order is shown in the inset; (b)-(d) absorption of three samples denoted as Design III (b), Design IV (c), and Design V (d), respectively. See Table 1 for resonance frequencies of the UVMA units comprising these samples. The blue scatters denote the distributions of the resonance frequencies.

## 2.5. Demonstration of High-Efficiency Air Ventilation

As mentioned in Section 2.4 (where the setup of the ventilation measurement are specified), the ventilation performance of the UVMA units are characterized by their wind velocity ratios [32,45,46], defined as the ratios between air flow velocities with and without the samples. First, Design I is characterized (open area ratio 72.4%), which is sandwiched between the two aluminum tubes. The electric fan is placed at the tube inlet and the anemometer is placed at the outlet, as depicted in Fig. 5(a). The measured air flow velocities (black dots) with ($v_{air,w}$) and without ($v_{air,wo}$) the sample are shown in Fig. 5(b). It can be seen that they manifest a linear dependence, and the linear fitting (red line) gives a favorable wind velocity ratio (*VR*) of 81.8%. Likewise, the wind velocity ratio of Design II (open area ratio 69.3%) is measured to be *VR* = 76.6%, as shown in Fig. 5(c). A direct demonstration of its ventilation performance is also provided (see Movie S4 in Supplementary Material). Thus, it proves that the design has achieved a high-efficiency ventilation while maintaining a near-unity low frequency sound absorption. Further, the broadband Design III is considered (open area ratio 69.3%), and its measured flow velocities are shown in Fig. 5(d). In a similar manner, the linear fitting gives a favorable wind velocity ratio (*VR*) of 62.3%. Although the broadband UVMA decreases the open area ratio to some extent, it can still maintain a desirable wind environment.

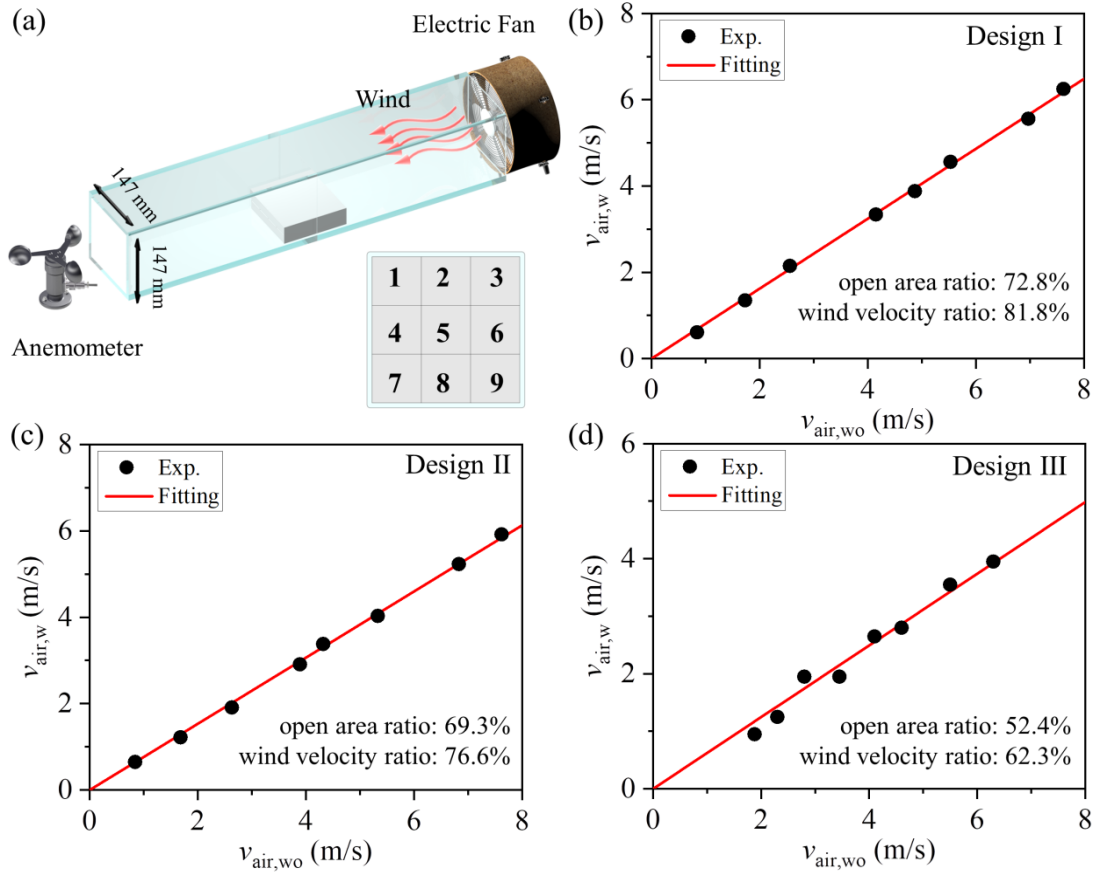

Fig. 5. (a) Schematic of the ventilation characterization system, which measures wind velocity ratios. The inset indicates the nine positions at the tube outlet where the air flow velocities are recorded and averaged. (c),(d) Averaged flow velocities at the outlet with ($v_{air,w}$) and without ($v_{air,wo}$) the samples, where the measured values (black dots) exhibit a good linear relation and can be well fitted (solid lines). The slopes of the fitted lines give the wind velocity ratios (*VR*) of Design I (c) and Design III (d). As denoted in the figures, *VR* = 81.8% for Design I, *VR* = 76.6% for Design II, and *VR* = 62.3% for Design III.

## 3. Discussion

An ultra-open ventilated metamaterial absorber with customized broadband performance working at low frequencies is proposed and demonstrated. The absorber is comprised of weakly coupled split-tube resonators, which can simultaneously achieve high-efficiency absorption and ventilation with proper geometric parameters (see Movies S1-S5 in Supplementary Material, which directly demonstrate its absorption and ventilation performances). The key to the absorption of the UVMA units is the small coupling between the two split-tube resonators, which leads to the merge of the resonance peaks of the symmetric and anti-symmetric modes. Due to the persistent high-efficiency absorption and the sensitive shift of the resonance frequency, the geometric parameters can be adjusted to achieve the required absorption performance. Through a combination of exponentially distributed resonance modes, the UVMAs can efficiently absorb sounds in a broadband range while still maintain a good ventilation. The absorber can also work in multiple discontinuous bands. This structure breaks the limits of previous acoustic absorbing metamaterials, achieving both high-efficiency acoustic absorption and ventilation, and can be extended to customized broadband working frequencies.

Therefore, the UVMA units should have promising application potential in a variety of acoustic engineering scenarios, such as the noise control of turbines and duct systems. Although the UVMA units immersed in air is the only situation investigated in this work, the design principle should also work for other background fluids, such as water.

Acknowledgements

X. Wu would like to acknowledge Prof. Lixi Huang for the fruitful discussion. This work was supported by the Hong Kong Research Grants Council (AoE/P-02/12 and 16204019), Fundamental Research Funds for the Central Universities (2019CDYGYB017), National Natural Science Foundation of China (11974067), and Natural Science Foundation Project of CQ CSTC (cstc2019jcyj-msxmX0145).

Author Contributions

X.W. and W.W. conceived the original idea. X.X., X.L., and X.W. performed the simulations. X.W. derived the theory. Y.H. and S.W. supported fabrication process of the samples. X.X. carried out the experiments. X.L., P.W., H.H. and Q.M. helped in the experiment setup. X.X. and X.W. analyzed the data, prepared the figures, and wrote the manuscript. X.W., Y.H., S.W. and W.W. supervised the project. All authors contributed to scientific discussions of the manuscript.

Competing Interests

The authors declare no competing financial interests.

Supplementary Material

See Supplementary Material at [inserted URL] for the demonstration videos, which directly demonstrate absorption and ventilation performance of the UVMA absorbers, and also its comparison with commercial melamine foams.

**Movie S1.** Demonstration of absorption performance of the UVMAs. We use two samples of Design II in the demonstration. The sound is at 472 Hz, the absorption peak of the UVMAs. The loudspeaker is placed in the middle of the aluminum tube. After we insert the two samples, the sound level decreases more than 20 dB.

**Movie S2.** Demonstration of absorption performance of the reference samples, which are foams (Basotect G+, BASF, Germany) of the same dimensions with Design II (that is, Foam II in Fig. 3(b)). The sound is also at 472 Hz, and the loudspeaker is still placed in the middle of the aluminum tube. However, the foams can hardly absorb any sound, even after sealing the tubes, as the sound level only decreases 3 dB at most.

**Movie S3.** Demonstration of customized broadband absorption of the UVMA. We use two samples of Design V in the demonstration. The first chirp sound is between 486-545 Hz, and the second chirp sound is between 659-710 Hz, both are efficiently absorbed. The sound level generally decreases 12 dB.

**Movie S4.** Demonstration of ventilation performance of the UVMAs. As shown in the video, after we insert a Design II into the tube, the wind velocity is slightly decreased

from ~7.2 m/s to ~5.9 m/s.

**Movie S5.** Front inside view of the ventilation characterization system, which reveals its setup inside the aluminum tube. A Design II is placed in the middle of the tube. The wind is generated by the electric fan placed at the inlet of the tube.

**Appendix A. Absorption spectra plotted in dB scale**

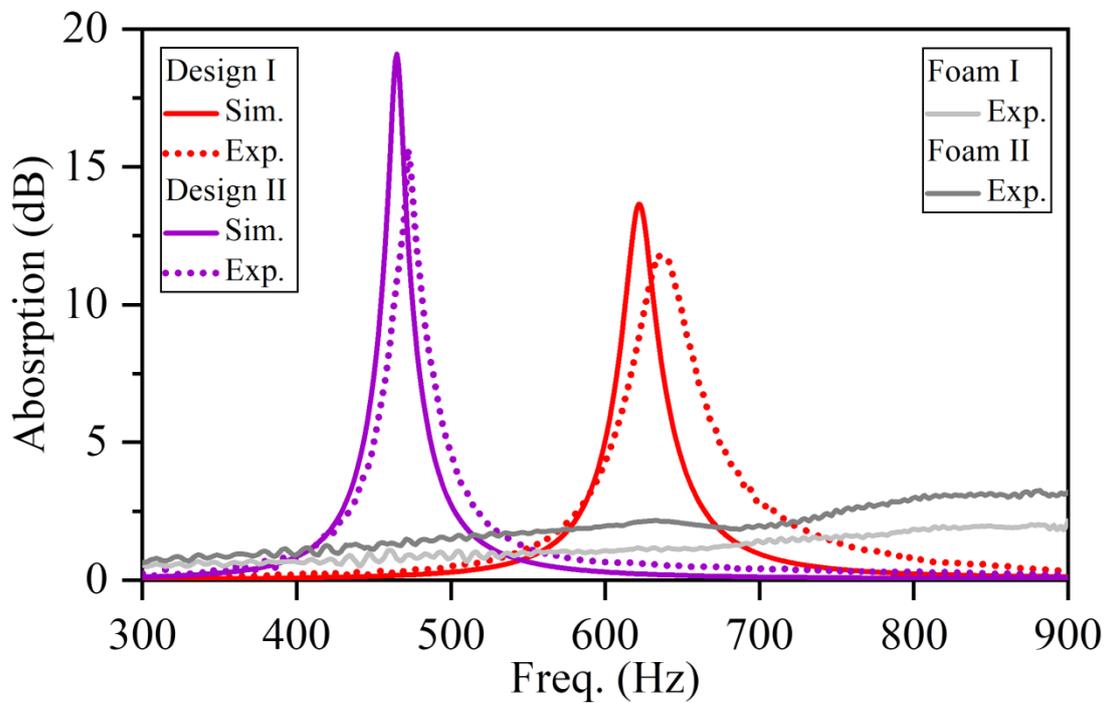

Fig. A.1. Simulated (colored solid lines) and experimentally measured (dotted lines) absorption spectra of Design I and Design II, plotted in dB scale. Gray solid lines represent measured absorption spectra of melamine foams (Basotect G+, BASF, Germany), also plotted in dB scale. As a reference sample, Foam I (II) has the same dimensions as Design I (II).

**Appendix B. Effective model of coupled lossy oscillators**

To analyze the interaction between the two resonators, we develop an effective model following previous models of coupled oscillators [31]. In brief, we model the UVMA as two coupled lossy oscillators, each with the mass $m$ and spring constant $k$, as shown in Fig. B.1(a). They also have identical visco-thermal loss and radiation loss. The motion of the two oscillators under external forces is then described by the equation

$$m\frac{d^2}{dt^2}\begin{bmatrix}x_1\\x_2\end{bmatrix}+\begin{bmatrix}\eta_l+\eta_r & \eta_c\\\eta_c & \eta_l+\eta_r\end{bmatrix}\frac{d}{dt}\begin{bmatrix}x_1\\x_2\end{bmatrix}+k\begin{bmatrix}x_1\\x_2\end{bmatrix}=\begin{bmatrix}F_1\\F_2\end{bmatrix}, \quad (B1)$$

where the subscripts denote each resonator (1, input side; 2, output side), $x_i$ representing the oscillation amplitude, $\eta_l$ representing the visco-thermal loss, $\eta_r$ representing the radiation loss, $\eta_c$ representing the radiation coupling between the two resonators, and $F_i$ is the force acting on the $i$-th oscillator which models the corresponding resonator. It can model a single pair of the coupled resonators in free space, or a unit of the coupled resonators arranged in an array, where the period effect results in different values of $\eta_r$ and $\eta_c$ compared with free space. For further analysis, we assume the $e^{-i\omega t}$ time-harmonic convention, which transforms the equation into frequency domain

$$\left\{-m\omega^2-i\omega\begin{bmatrix}\eta_l+\eta_r & \eta_c\\\eta_c & \eta_l+\eta_r\end{bmatrix}+k\right\}\begin{bmatrix}x_1\\x_2\end{bmatrix}=\begin{bmatrix}F_1\\F_2\end{bmatrix}. \quad (B2)$$

With the vibration amplitude ratio $X_{ij}$ defined as $X_{ij}=x_i/x_j$, we have

$$\left\{-m\omega^2-i\omega\begin{bmatrix}\eta_l+\eta_r+\eta_c X_{21} & \\ & \eta_l+\eta_r+\eta_c X_{12}\end{bmatrix}+k\right\}\begin{bmatrix}x_1\\x_2\end{bmatrix}=\begin{bmatrix}F_1\\F_2\end{bmatrix}. \quad (B3)$$

Thus, we obtain the expressions for $x_1$ and $x_2$

$$x_1 = \frac{F_1}{-m\omega^2 - i\omega(\eta_l + \eta_r + \eta_c X_{21}) + k},$$
$$x_2 = \frac{F_2}{-m\omega^2 - i\omega(\eta_l + \eta_r + \eta_c X_{12}) + k}$$
(B4)

and we then have

$$X_{21} = x_2/x_1 = \frac{-m\omega^2 - i\omega(\eta_l + \eta_r + \eta_c X_{21}) + k}{-m\omega^2 - i\omega(\eta_l + \eta_r + \eta_c X_{12}) + k} F_{21},$$
(B5)

where the force ratio $F_{ij} = F_i/F_j$. Note that $X_{12} = 1/X_{21}$, solving Eq. (B5) leads to an explicit formula for $X_{21}$

$$X_{21} = \frac{2\eta_c - 2F_{21}(\eta_l + \eta_r) + iF_{21}\eta_0(\frac{\omega}{\omega_0} - \frac{\omega_0}{\omega})}{2(F_{21}\eta_c - \eta_l - \eta_r) + i\eta_0(\frac{\omega}{\omega_0} - \frac{\omega_0}{\omega})},$$
(B6)

in which the parameters $\omega_0 = \sqrt{k/m}$ is the resonance frequency and $\eta_0 = 2\sqrt{km}$ is the critical damping coefficient. The dissipation power of the oscillators, averaged over a period, is then [31]

$$P_{diss} = \frac{1}{2}\text{Re}\left[\eta_l \frac{dx_1^*}{dt}\frac{dx_1}{dt}\right] + \frac{1}{2}\text{Re}\left[\eta_l \frac{dx_2^*}{dt}\frac{dx_2}{dt}\right]$$
$$= \frac{1}{2}\omega^2 \eta_l |x_1|^2 (1 + |X_{21}|^2)$$
$$= \frac{2\eta_l |F_1|^2 (1 + |X_{21}|^2)}{\eta_0^2[\frac{\omega_0}{\omega} - \frac{\omega}{\omega_0} + 2\frac{\text{Im}(\eta_r + \eta_c X_{21})}{\eta_0}]^2 + 4[\eta_l + \text{Re}(\eta_r + \eta_c X_{21})]^2}.$$
(B7)

For one-sided incidence, it can be viewed as an linear superposition of symmetric and anti-symmetric incidences [23], hence we express the external forces as $F_1 = pS_0$ and $F_2 = 0$, respectively, where $S_0$ is the effective area of the resonator, and $p$ is the incident acoustic pressure. We have the force ratio $F_{21} = 0$, and the dissipation power

is then

$$P_{diss} = \frac{2\eta_l |p|^2 S_0^2 (1+|X_{21,0}|^2)}{\eta_0^2 [\frac{\omega_0}{\omega} - \frac{\omega}{\omega_0} + 2\frac{\text{Im}(\eta_r + \eta_c X_{21,0})}{\eta_0}]^2 + 4[\eta_l + \text{Re}(\eta_r + \eta_c X_{21,0})]^2}, \quad (B8)$$

with

$$X_{21,0} = X_{21}|_{F_{21}=0} = \frac{2\eta_c}{-2(\eta_l + \eta_r) + i\eta_0(\frac{\omega}{\omega_0} - \frac{\omega_0}{\omega})} \quad (B9)$$

On the other hand, the incident power can be expressed as

$$P_{inc} = \frac{|p|^2 S_{inc}}{2Z}, \quad (B10)$$

where $S_{inc}$ is the area of the incidence channel, and $Z$ is the specific acoustic impedance of air. Therefore, The absorption coefficient of the oscillators, equivalent to the ratio between dissipation power and incident power is then

$$A = P_{diss} / P_{inc}$$
$$= \frac{2\eta_l \eta_r^0 (1+|X_{21,0}|^2)}{\eta_0^2 [\frac{\omega_0}{\omega} - \frac{\omega}{\omega_0} + 2\frac{\text{Im}(\eta_r + \eta_c X_{21,0})}{\eta_0}]^2 + 4[\eta_l + \text{Re}(\eta_r + \eta_c X_{21,0})]^2}, \quad (B11)$$

where $\eta_r^0 = ZS_{inc}/S_0^2$ is an auxiliary parameter, which we refer to as the reference radiation loss.

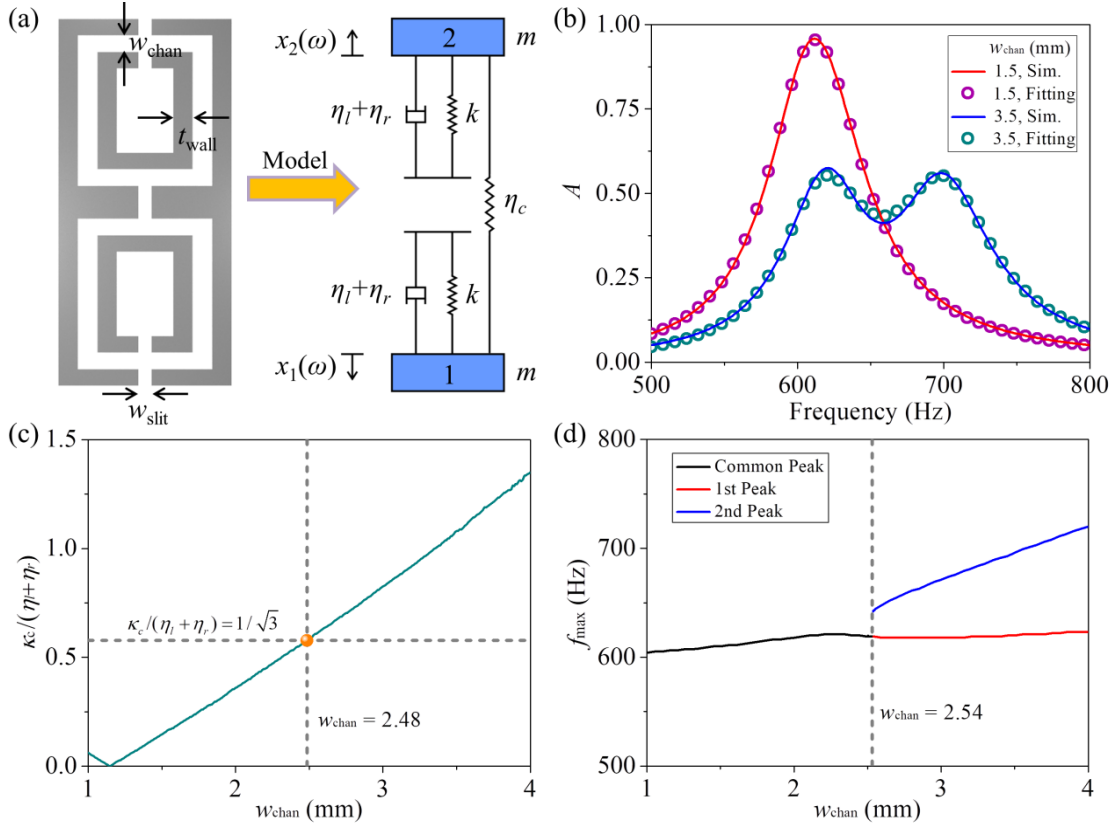

Fig. B.1. (a) Cross section (*xz*-plane) of the UVMA unit, and its effective model comprised of lossy mass and springs. The subscripts denote each resonator (1, input side; 2, output side). (b) Simulated (solid lines) and fitted (scatters) absorption spectra of the UVMA unit using Eq. (B11) when $w_{chan}$ =1.5 mm and $w_{chan}$ = 3.5 mm, respectively. (c) The ratio $\kappa_c/(\eta_r+\eta_l)$ retrieved from simulations. (d) The frequency of absorption peaks retrieved from simulations.

Obviously, the maximum or minimum absorptions happen at the frequency satisfying the condition

$$\frac{\partial A}{\partial \omega} = 0 . \quad (B12)$$

For possibility of an analytic analysis to obtain intuitive insight, we consider the deep-subwavelength case, such that $\eta_r$ is pure real and $\eta_c = i\kappa_c$ is pure imaginary. After cumbersome algebra, it is found that Eq. (B12) has a direct solution

$$\omega = \omega_0, \tag{B13}$$

and the other solutions are determined by the equation

$$\eta_0^4 (\frac{\omega}{\omega_0} - \frac{\omega_0}{\omega})^4 + 8\eta_0^2[(\eta_l + \eta_r)^2 + \kappa_c^2](\frac{\omega}{\omega_0} - \frac{\omega_0}{\omega})^2 \\ + 16[(\eta_l + \eta_r)^2 - 3\kappa_c^2][(\eta_l + \eta_r)^2 + \kappa_c^2] = 0 \tag{B14}$$

In fact, when Eq. (B14) has real solutions, the absorption spectrum will exhibit two peaks around the dip at $\omega_0$. The existence of the real solutions leads to the inequality

$$16[(\eta_l + \eta_r)^2 - 3\kappa_c^2][(\eta_l + \eta_r)^2 + \kappa_c^2] < 0, \tag{B15}$$

that is,

$$\kappa_c > \frac{\eta_l + \eta_r}{\sqrt{3}}. \tag{B16}$$

The solutions are then

$$\frac{\omega}{\omega_0} - \frac{\omega_0}{\omega} = \frac{2}{\eta_0}\sqrt{2\kappa_c\sqrt{(\eta_l + \eta_r)^2 + \kappa_c^2} - [(\eta_l + \eta_r)^2 + \kappa_c^2]}, \tag{B17}$$

which, after some algebra, gives

$$\omega_\pm = \omega_0(\chi \pm \sqrt{1 + \chi^2}), \tag{B18}$$

where

$$\chi = \frac{1}{\eta_0}\sqrt{2\kappa_c\sqrt{(\eta_l + \eta_r)^2 + \kappa_c^2} - [(\eta_l + \eta_r)^2 + \kappa_c^2]}. \tag{B19}$$

The maximum absorption $A_{\max}$ is then

$$A_{\max} = A(\omega_\pm) = \frac{\eta_l \eta_r^0 [\kappa_c + \sqrt{(\eta_l + \eta_r)^2 + \kappa_c^2}]}{8(\eta_l + \eta_r)^2 \kappa_c}. \tag{B20}$$

Eqs. (B16) and (B20) confirms that the split of maximum absorptions observed in Fig.

3(d) is due to the large coupling radiation between the resonators.

On the other side, if the radiation coupling $\kappa_c$ becomes small enough, such that

$$\kappa_c \leq \frac{\eta_l + \eta_r}{\sqrt{3}}, \tag{B21}$$

the two absorption peaks will merge together since Eq. (B14) now has no real solutions. Correspondingly, the maximum absorption $A_{\max}$ will happen at $\omega_0$, with the value

$$A_{\max} = A(\omega_0) = \frac{\eta_l \eta_r^0}{2[(\eta_l + \eta_r)^2 + \kappa_c^2]}. \tag{B22}$$

We then retrieve the parameters from COMSOL Multiphysics to confirm our above discussion. More specifically, the parameters $\kappa_c$ is determined by the frequency split between symmetric and anti-symmetric modes, while $(\eta_r+\eta_l)$ is determined by the imaginary part of the eigenmodes. Other parameters are retrieved from fitting of the simulated absorption spectra using Eq. (B11). We consider two cases, $w_{\text{chan}}$ = 1.5 mm and $w_{\text{chan}}$ = 3.5 mm, respectively. The fitted absorption spectra are plotted in Fig. B.1(b), and they agree excellently with simulation results in both cases. Further, the ratio $\kappa_c/(\eta_r+\eta_l)$ is plotted in Fig. B.1(c), and it crosses the critical value $1/\sqrt{3}$ at $w_{\text{chan}}$ = 2.48 mm, as denoted by the dashed lines. On the other side, the merge of absorption peaks happens at $w_{\text{chan}}$ = 2.54 mm, as summarized in Fig. B.1(d), which generally agrees with the value predicted form our effective model. The small error (2.4%) between the two values could be attributed to the fact that we do not consider the phase of the radiation coupling ($\eta_c$ assumed to be pure imaginary) between the two resonators.

**Appendix C. Emergence of 90°-phase difference with coalesced absorption peaks**

As mentioned above, When the two absorption peaks are merged, the maximum absorption happens at $\omega_0$. For (angular) frequency around $\omega_0$, we denote $\omega = \omega_0 + \delta\omega$, and the vibration ratio for one-sided incidence can be expanded as

$$X_{21,0}(\omega_0 + \delta\omega) = \frac{2i\kappa_c}{-2(\eta_l + \eta_r) + i\eta_0(\dfrac{\omega_0 + \delta\omega}{\omega_0} - \dfrac{\omega_0 + \delta\omega}{\omega})} \\ \approx -\frac{i\kappa_c}{\eta_l + \eta_r} + \frac{\eta_0 \kappa_c}{(\eta_l + \eta_r)^2 \omega_0}\delta\omega \quad (C1)$$

where we have neglected higher order terms. The phase of $X_{21,0}$ at resonance ($\delta\omega = 0$) is then exactly −90°. Since the 1st resonator is directly impinged by the incident sound, the phase of the 2nd resonator lags behind the 1st resonator, as expected.

**Appendix D. Absorption spectra when terminated by different boundaries**

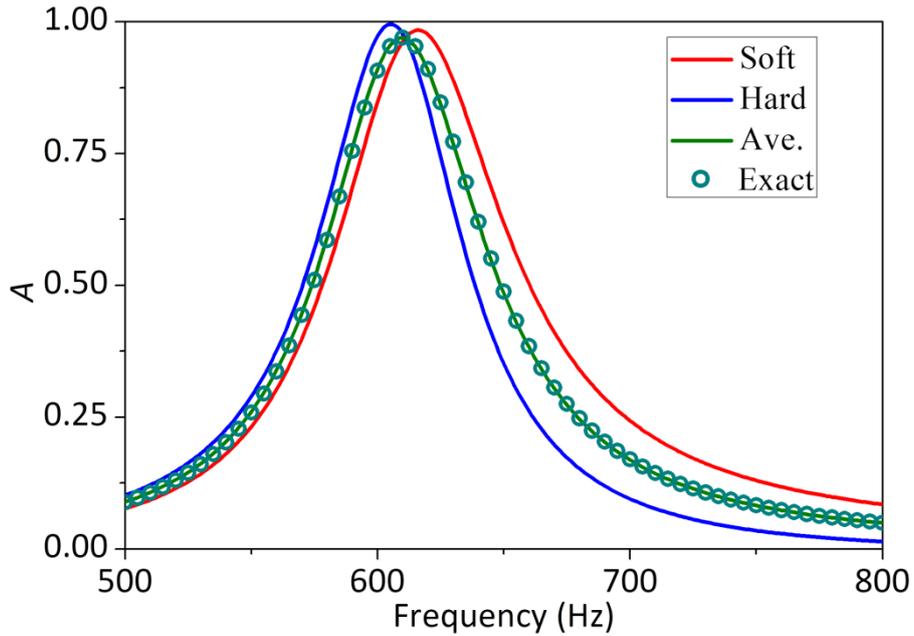

Fig. D.1. Calculated absorption spectra when considering the situation of a single

split-tube resonator terminated by an acoustic soft boundary (red solid line) or an acoustic hard boundary (blue solid line). Their average (green solid line) agrees excellently with the exact situation (green scatters), which confirms that the mirror plane ($z = 0$ denoted in Fig. 3(c)) can be viewed as an superposition of the acoustic soft and hard boundaries.

## Appendix E. UVMAs under oblique incidence

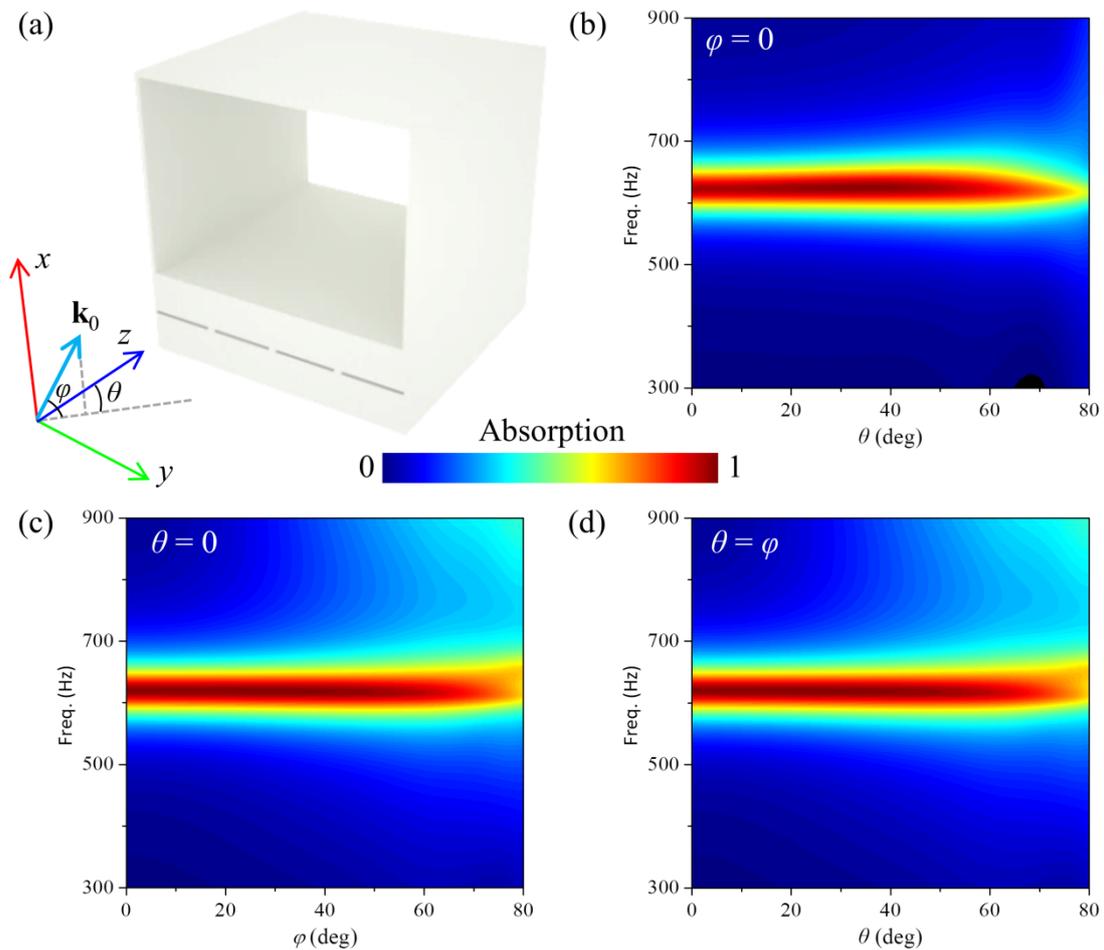

Fig. E.1. Simulated absorptions under oblique incidence. (a) Definition of the angles of incidence, $\theta$ and $\varphi$. $\mathbf{k}_0$ represents the wave vector of incident sound. (b)-(d) Simulated spectra of absorption as the function of frequency and the angles of incidence, in the direction $\varphi = 0$ (b), $\theta = 0$ (c), and $\theta = \varphi$ (d). Here the array is comprised of Design I (open area ratio 72.8%), with geometric parameters $a = 100$ mm, $b = 40$ mm, $w_{chan} = 1.4$ mm, $w_{slit} = 1.4$ mm.

**Appendix F. Photographs of experiment setups**

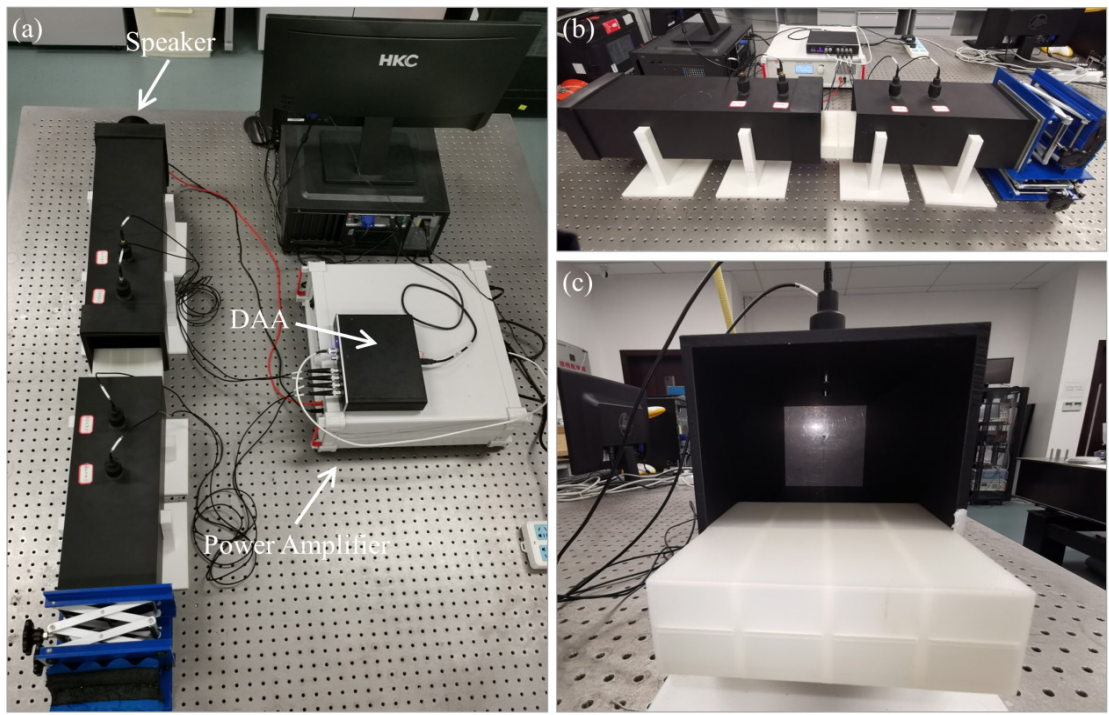

Fig. F.1. (a)-(c) Photographs of the square cross-section impedance tube. (a) Top view of the impedance tube. The speaker, power amplifier, and data acquisition analyzer (DAA) are denoted by the arrows. (b) Side view of the setup. (c) Front view of the tube loaded with a sample.

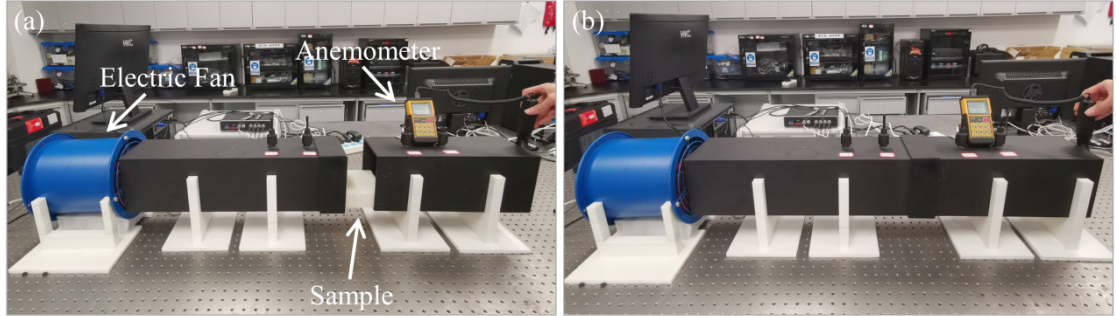

Fig. F.2. (a),(b) Photographs of the ventilation characterization system. The sample, electric fan, and anemometer are denoted by the arrows. During measurement (b), the two aluminum tubes are sealed together.

**Appendix G. Materials and Methods**

**G.1. Fabrication of UMVA Units**

All UVMA units are fabricated using stereolithography (SLA) 3D printing techniques. The 3D printer (iSLA660, ZRapid Tech, China) uses ultraviolet (UV) light to cure liquid photopolymer resin layer by layer in the fabrications. The 3D printer has an accuracy of 0.2 mm. The UV-cured resin has a tensile modulus of 2.46 GPa and a density of 1.10 g/cm$^3$.

**G.2. Setup of Simulations**

All full-wave simulations are performed using COMSOL Multiphysics, a commercial finite element method (FEM) solver. Coupled modules of Pressure Acoustics and Thermoacoustics are used. The thermal and viscous losses inside the UVMA units are naturally included in Thermoacoustics module. In all simulations, the 3D-printed photopolymer resins are treated as acoustic hard boundaries, while the material parameters of air are set as follows: density 1.2 kg/m$^3$, dynamic viscosity 18.5 μPa, thermal conductivity 24 mW/(m·K), specific heat ratio 1.4, thermal expansion coefficient $3.41 \times 10^{-3}$ K$^{-1}$, and speed of sound 343 m/s. No tunable auxiliary parameters exist in the simulations.

### G.3. Setup of Impedance Tube

All acoustic measurements are performed in a square impedance tube using the general four-microphone two-load method [23,47]. The impedance tube (schematically depicted in Fig. 1(e)) is comprised of two aluminum square tubes (inner cross section of 147×147 mm$^2$, tube thickness 5 mm), a full-range speaker (M5N, HiVi, China), four microphones (MP418, BSWA, China), a power amplifier (ATA 304, Aigtek, China), and a data acquisition analyzer (MC3242, BSWA, China). The plane wave cutoff frequency of the aluminum tubes is ~1100 Hz. The lengths of the two aluminum tubes are 600 mm and 400 mm, respectively. A clamped aluminum plate of thickness 4 mm is employed as the rigid back to model acoustic hard-boundary termination. After removing the aluminum plate, the sound in the tube will radiate outside that models an acoustic open boundary termination. They serve as two different termination loads in the measurement. Photographs of the impedance tube are shown in Fig. F.1 (see Appendix F).

### G.4. Setup of Ventilation Measurement

Ventilation efficiency of the UVMA units are characterized using a measurement system [32] comprised of two aluminum square tubes (inner cross section of 147×147 mm$^2$, tube thickness of 5 mm) with a length of 200 mm and an electric fan (SF2-2, HEYUNCN, China) with a maximum air volume of $3.7 \times 10^3$ m$^3$/h. An anemometer (TM856, TECMAN, China) is used to monitor the air flow velocities at the outlet of the aluminum tube, while the electric fan is placed at the inlet, as shown in Fig. 5(a). The air gap between the electric fan and the tube are sealed with sponges. We use the wind velocity ratio (*VR*) to characterize the ventilation performance defined as the ratio between the measured air velocities with and without the sample (*VR* = $v_{air,w}$ / $v_{air,wo}$). Here, $v_{air,w}$ and $v_{air,wo}$ refers to the average flow velocity at the outlet with/without the sample placed between the aluminum tubes, respectively. Moreover, to improve the measurement reliability, the air flow velocities are recorded at the positions 1–9 of the outlet (as shown in the inset of Fig. 5(a)). The readings on the anemometer are averaged to obtain the final data. The inlet air flow velocities are varied by tuning the power of the electric fan, such that different pairs ($v_{air,w}$, $v_{air,wo}$) of the averaged air flow velocities are obtained and are used to calculate the wind velocity ratio (*VR*). Photographs of the ventilation measurement system are shown in Fig. F.2 (see Appendix F). A direct view of the measurement system inside the tube is also given in Movie S5 (see Supplementary Material).